\documentclass{article}
\usepackage{times}
\usepackage{graphics}

\def\jpsi{\mbox{$\mathrm{J/}\Psi$}}

\def\sqrtsNN{\mbox{$\sqrt{s_\mathrm{NN}}$}}

\def\TsubA{\mbox{$\mathrm{T}_\mathrm{A}$}}
\def\TsubB{\mbox{$\mathrm{T}_\mathrm{B}$}}
\def\TsubAB{\mbox{$\mathrm{T}_\mathrm{AB}$}}

\def\RsubA{\mbox{$\mathrm{R}_\mathrm{A}$}}
\def\bsubB{\mbox{$\mathrm{b}_\mathrm{B}$}}
\def\bsubA{\mbox{$\mathrm{b}_\mathrm{A}$}}

\def\sigmaNN{\mbox{$\sigma_\mathrm{NN}$}}
\def\sigmaNNhard{\mbox{$\sigma_\mathrm{NN}^\mathrm{hard}$}}

\def\vone{\mbox{$\mathrm{v}_1$}}
\def\vtwo{\mbox{$\mathrm{v}_2$}}

\def\absvecs{\mbox{$|\vec{\mathrm{s}}|$}}

\def\smean{\mbox{$<\!\absvecs\!>$}}

\begin{document}

\title{Spatial Distribution of Initial Interactions in High Energy Collisions of Heavy Nuclei}
\author{
Peter Jacobs\thanks{email: pmjacobs@lbl.gov}{\ } and
Glenn Cooper\thanks{email: glenn@qedinc.com}\\
Lawrence Berkeley National Laboratory\\
Berkeley, CA 94720
}
\date{
August 23, 2000
}

\maketitle

\begin{abstract}

The spatial distribution of interactions in high energy collisions of
heavy nuclei is discussed using the wounded nucleon, binary collision,
hard sphere, and colliding disk parameterizations of interaction
densities. The mean radius, its dispersion, and the eccentricity of
the interaction region are calculated as a function of impact
parameter. The eccentricity is of special interest for comparison to
measurements of anisotropic flow. The number of participants and
binary collisions is also tabulated as a function of impact parameter.

\end{abstract}

\section {Introduction}

In this note we discuss the spatial distribution of initial
interactions in the high energy collisions of heavy nuclei, in order
to gain some insight into geometrical effects in such collisions. We
apply a simple, widely used formalism~(e.g. \cite{kari,ramona}) that
is often attributed to Glauber~\cite{glauber}, to calculate the
density of interactions projected onto the plane transverse to the
beam axis for four commonly used weighting functions: number of
wounded nucleons, number of binary collisions, colliding hard spheres,
and colliding disks. The wounded nucleon and binary collision
calculations incorporate the usual Woods-Saxon parametrization of
nuclear density. We compute the mean and dispersion of the radius and
the eccentricity of the interaction zone as a function of impact
parameter for symmetric collisions of heavy nuclei. Using the same
calculation we tabulate the number of participants for the wounded
nucleon and hard sphere paramterizations, together with the number of
binary collisions, as a function of impact parameter.

The distributions calculated are implicit in all model calculations of
low or high sophistication that incorporate similar parameterizations
of nuclear geometry, and in that sense nothing new is presented
here. However, we have not found these distributions collected and
discussed explicitly in the literature, and that is the the purpose of
this note. A more detailed discussion of these and related results was
presented previously in~\cite{sn0402}.

The formalism is described briefly in section 2. Section 3 discusses the
transverse profile of the interaction volume as a function of impact
parameter. Section 4 presents a parametrization of the eccentricity of
the interaction region for non-central collisions. Section 5 tabulates
the number of participants for the wounded nucleon and hard sphere
paramterizations, together with the number of binary collisons, as a
function of impact parameter.

\section {Formalism}

\begin{figure}
\resizebox{4in}{!}{\includegraphics*[0in,0in][5in,3.5in]{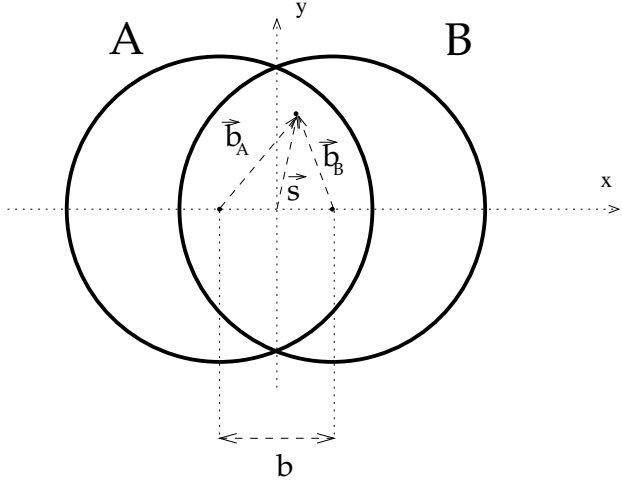}}
\caption{Transverse coordinate system. A and B are the masses of the
  colliding nuclei. The incoming beams are (anti-)parallel to the z
  axis. The x-z plane is commonly called the reaction plane.}
\label{fig:coordsystem}
\end{figure}

The coordinate system is defined in Figure~\ref{fig:coordsystem}. We
utilize the standard nuclear thickness function \TsubA~\cite{ramona},
\begin{equation}
\TsubA(|\vec{s}|) = \int{dz}\rho_\mathrm{A}(z,\vec{s}). 
\label{eq:TsubA}
\end{equation}
For the nuclear density we use a Woods-Saxon distribution,
\begin{equation}
\rho_\mathrm{A}(r)=\rho_0\cdot\frac{1}{1+e^{(r-\RsubA)/\mathrm{a}}},
\label{eq:woodssaxon}
\end{equation}
where $r=\sqrt{s^2+z^2}$, $\RsubA=1.12\cdot\mathrm{A}^{1/3}$, and
$\rho_0=0.159$ GeV/$\mathrm{fm}^3$ and a=0.535 fm for
$\ ^{197}\mathrm{Au}$.  $\rho_\mathrm{A}$ is normalized so that
$\int{d^2s}\TsubA(|\vec{s}|)=\mathrm{A}$.
 
In Fig.~\ref{fig:coordsystem}, the nuclear impact parameter is denoted
b, and the distance to a point in the transverse plane from the center
of either nucleus is denoted \bsubA\ and \bsubB. The vector from the
origin to this point is written
\begin{equation}
\vec{s}=\vec{\bsubA}-\frac{b}{2}\hat{x}=\vec{\bsubB}+\frac{b}{2}\hat{x}.
\label{eq:s}
\end{equation}
where $\hat{x}$ is the unit vector in the $x$ direction.

To calculate the density of interactions as a function of impact
parameter for a given process, we utilize the limiting cases of high
and low nucleon-nucleon cross section for that process:

\begin{itemize}
\item{Wounded Nucleon Scaling:} We consider the number of nucleons at
  $\vec{s}$ that are struck at least once by the nucleons in the
  oncoming nucleus, where ``struck'' means inelastically excited with
  nucleon-nucleon collision cross section\footnote{\sigmaNN$\approx$
    30 mb at the SPS (\sqrtsNN=20 GeV) and 40 mb at RHIC (\sqrtsNN=200
    GeV).} \sigmaNN. In the transverse projection, the density of
  wounded nucleons per nuclear collision is given in units
  $1/\mathrm{fm}^2$ by:
\begin{equation}
\frac{\mathrm{d^2N}_\mathrm{WN}}{\mathrm{d}\mathrm{s}^2}=
\TsubA(\bsubA)\cdot(1-e^{-\TsubB(\bsubB)\sigmaNN})+
\TsubB(\bsubB)\cdot(1-e^{-\TsubA(\bsubA)\sigmaNN}),
\label{eq:woundednucleon}
\end{equation}
where the dependence on nuclear impact parameter is via Eq.~\ref{eq:s}.

\item{Binary Collision Scaling:} More generally, we mean those
  processes with sufficiently small nucleon-nucleon cross section
  \sigmaNNhard\ that $(\TsubA\cdot\sigmaNNhard){\ll}1$ (we apply the
  label ``hard'' because we are usually refering to high momentum
  transfer processes, or hard scattering). Nucleon-nucleon interaction
  probabilities can then be summed for the total interaction probability,
  and the number of hard scatterings per nuclear collision goes as the
  number of binary nucleon-nucleon collisions. The density in the
  transverse projection of hard processes per nuclear collision, in
  units $1/\mathrm{fm}^2$, is then
\begin{equation}
\frac{\mathrm{d^2N}_\mathrm{hard}}{\mathrm{d}\mathrm{s}^2}=
\sigmaNNhard\cdot\TsubA(\bsubA)\cdot\TsubB(\bsubB).
\label{eq:hardprocess}
\end{equation}
Integrating over the transverse plane (and changing integration
variables), the total number of hard processes per nuclear collision
is given as a function of nuclear impact parameter b
by\cite{kari,ramona}:
\begin{equation}
\mathrm{N}_\mathrm{hard}(b)=\sigmaNNhard\cdot\int{d^2s}\TsubA(|\vec{s}|)
\TsubB(|\vec{b}-\vec{s}|)\equiv\sigmaNNhard\cdot\TsubAB(b)
\label{eq:hardprocessN}
\end{equation}

\end{itemize}

\section{Transverse Profile of the Interaction Volume for Non-central
  Collisions}

The azimuthal anisotropy of momentum distributions in nuclear
collisions can be related to the orientation of the event plane
(the x-z plane in Figure~\ref{fig:coordsystem}) for noncentral
collisions over a wide range of energies~\cite{sn388}, and has been
quantitatively studied up to the highest energy nuclear collisions at
the SPS~\cite{na49flow} and RHIC~\cite{starflow}. The lowest order harmonics are referred to
as directed and elliptic flow, and are characterized by coefficients
\vone\ and \vtwo\ respectively in a Fourier expansion of
the azimuthal angle or momentum distributions~\cite{sn388}.

Elliptic flow at midrapidity has received particular attention
recently because of a possible connection to the Equation of State
(see discussion and references in~\cite{sn388}). To help distinguish
dynamics from purely geometrical effects, it has been
suggested~\cite{sorgealpha,heiselberg} that the measured \vtwo, the
elliptic anisotropy, be scaled by the eccentricity of the
reaction volume. This is defined to be~\cite{sorgealpha,heiselberg}
\begin{equation}
\epsilon\equiv\frac{<y^2>-<x^2>}{<y^2>+<x^2>}\approx\frac{b}{2\RsubA}
\label{eq:epsilon}
\end{equation}
where $<\ldots>$ indicates the spatial average over the transverse
plane weighted by a density such as that of wounded nucleons
(equation~\ref{eq:woundednucleon}). The approximation is the ratio of
the lengths of the axes of the overlap region in
Figure~\ref{fig:coordsystem}, $(y\mid_{x=0})/(x\mid_{y=0})$, not
weighted by nuclear density~\cite{heiselberg}.

We calculate the transverse density distribution of the reaction
volume as a function of impact parameter, utilizing four different
weighting functions:
\begin{itemize}
\item{Wounded Nucleons:} The transverse density profile is calculated
  using a Woods-Saxon density distribution and
  Eq.~\ref{eq:woundednucleon}, appropriate for the bulk of particle
  production. We use \sigmaNN=30 mb, but also investigate the
  sensitivity of the computed quantities to this parameter.
\item{Binary Collisions:} The transverse density profile is calculated
  using a Woods-Saxon density distribution and
  Eq.~\ref{eq:hardprocess}, appropriate for hard processes such as jet
  and \jpsi\ production.
\item{Hard Sphere:} The transverse density profile is calculated for
  colliding sharp-edged spheres, with the density defined as
  \TsubA+\TsubB. This corresponds to the limit of the Wounded Nucleon
  density in which the Woods-Saxon parameter $a$ in
  Eq.~\ref{eq:woodssaxon} is small and \sigmaNN\ in
  Eq.~\ref{eq:woundednucleon} is large. The radius of the hard sphere
  corresponding to an Au nucleus is 7.24 fm, increased from the
  Woods-Saxon value of 6.52 fm in order to obtain the same total interaction cross
  section.
\item{Two Dimensional:} The transverse density profile is simply the
  area of overlap region in Fig.~\ref{fig:coordsystem}, with uniform
  weighting (i.e.\ without taking into account the nuclear thickness in
  the z-direction and corresponding variation in density when projected onto the plane). 
  The radius corresponding to a Au nucleus is 7.24
  fm, as in the Hard Sphere calculation.
\end{itemize}


\begin{figure}
\resizebox{\textwidth}{!}{\includegraphics{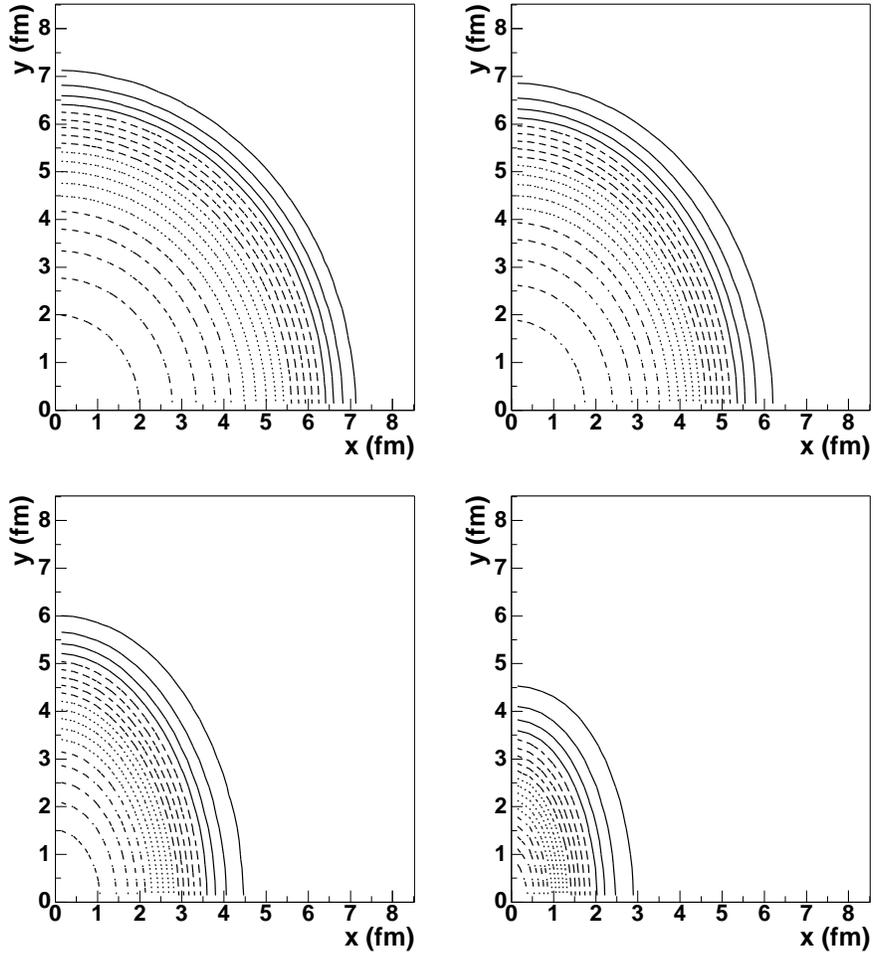}}
\caption{Transverse density of Wounded Nucleons as a function of (x,y) (see
  Fig~\ref{fig:coordsystem}), for collisions of Au+Au at impact
  parameters b=0, 4, 8 and 12 fm.}
\label{fig:densitywn}
\end{figure}

\begin{figure}
\resizebox{\textwidth}{!}{\includegraphics{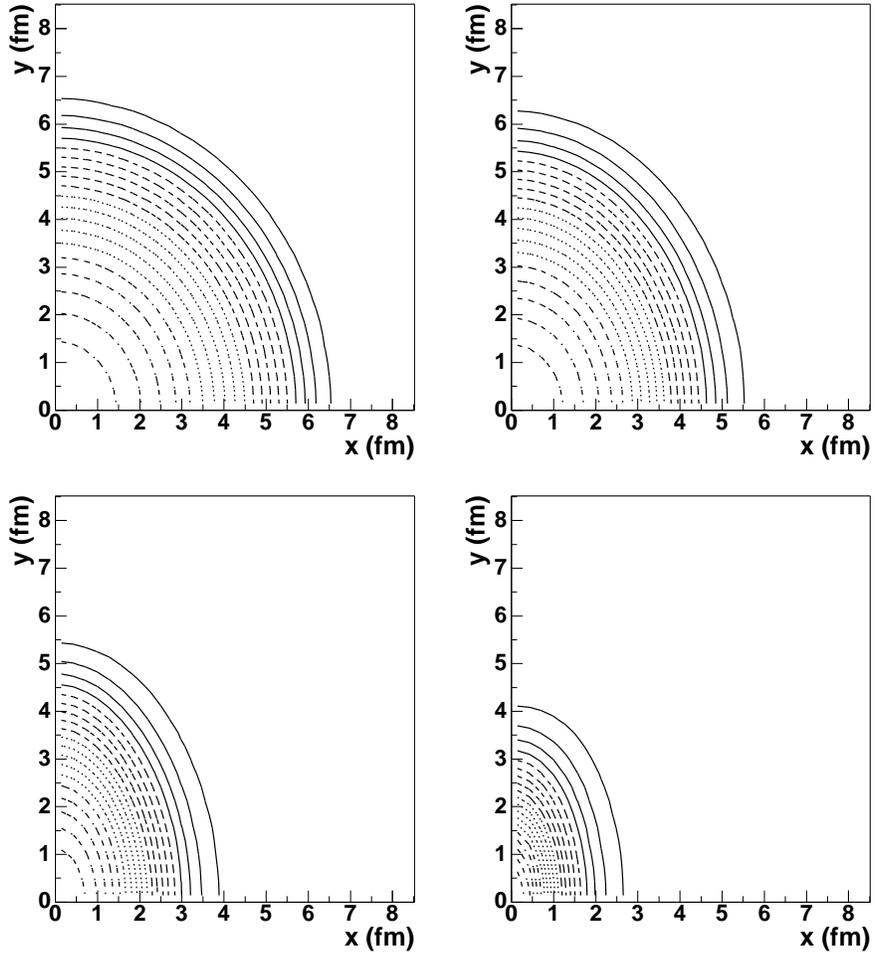}}
\caption{Transverse density of Binary Collisions as a function of (x,y) (see
  Fig~\ref{fig:coordsystem}), for collisions of Au+Au at impact
  parameters b=0, 4, 8 and 12 fm.}
\label{fig:densitybc}
\end{figure}

\begin{figure}
\resizebox{\textwidth}{!}{\includegraphics{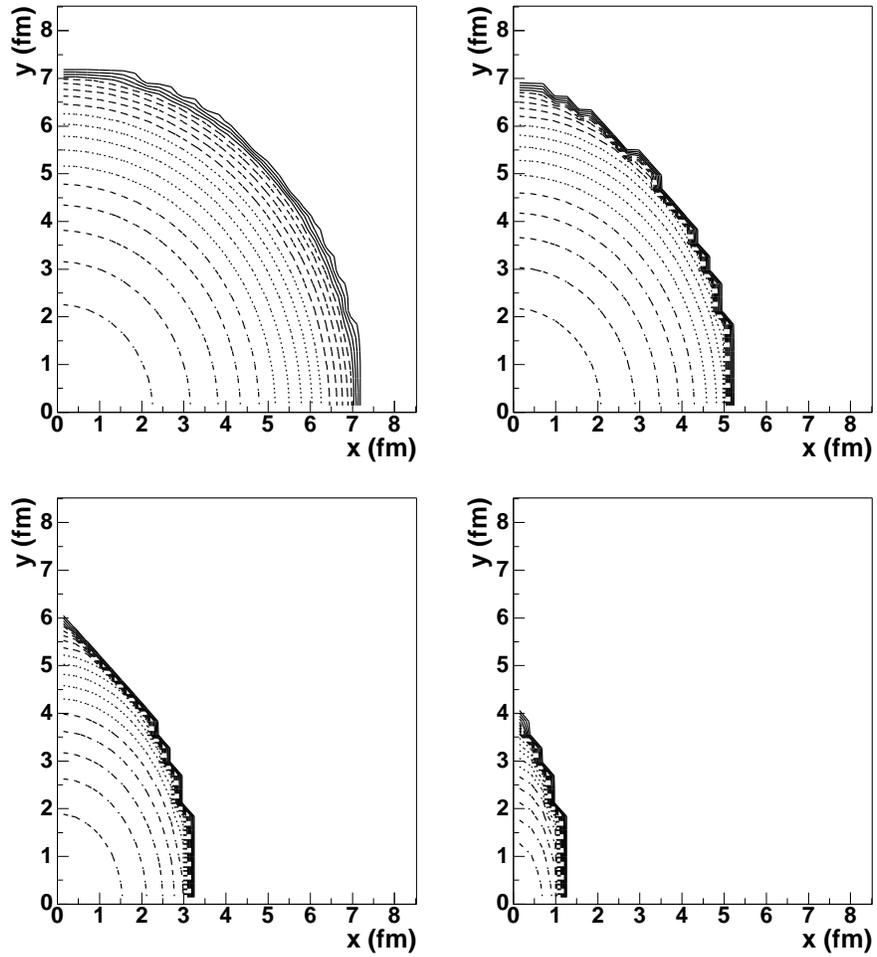}}
\caption{Transverse density profile of interacting hard spheres as 
  a function of (x,y) (see Fig~\ref{fig:coordsystem}), corresponding
  to Au+Au at impact parameters b=0, 4, 8 and 12 fm.}
\label{fig:densityhs}
\end{figure}

\begin{figure}
\resizebox{\textwidth}{!}{\includegraphics{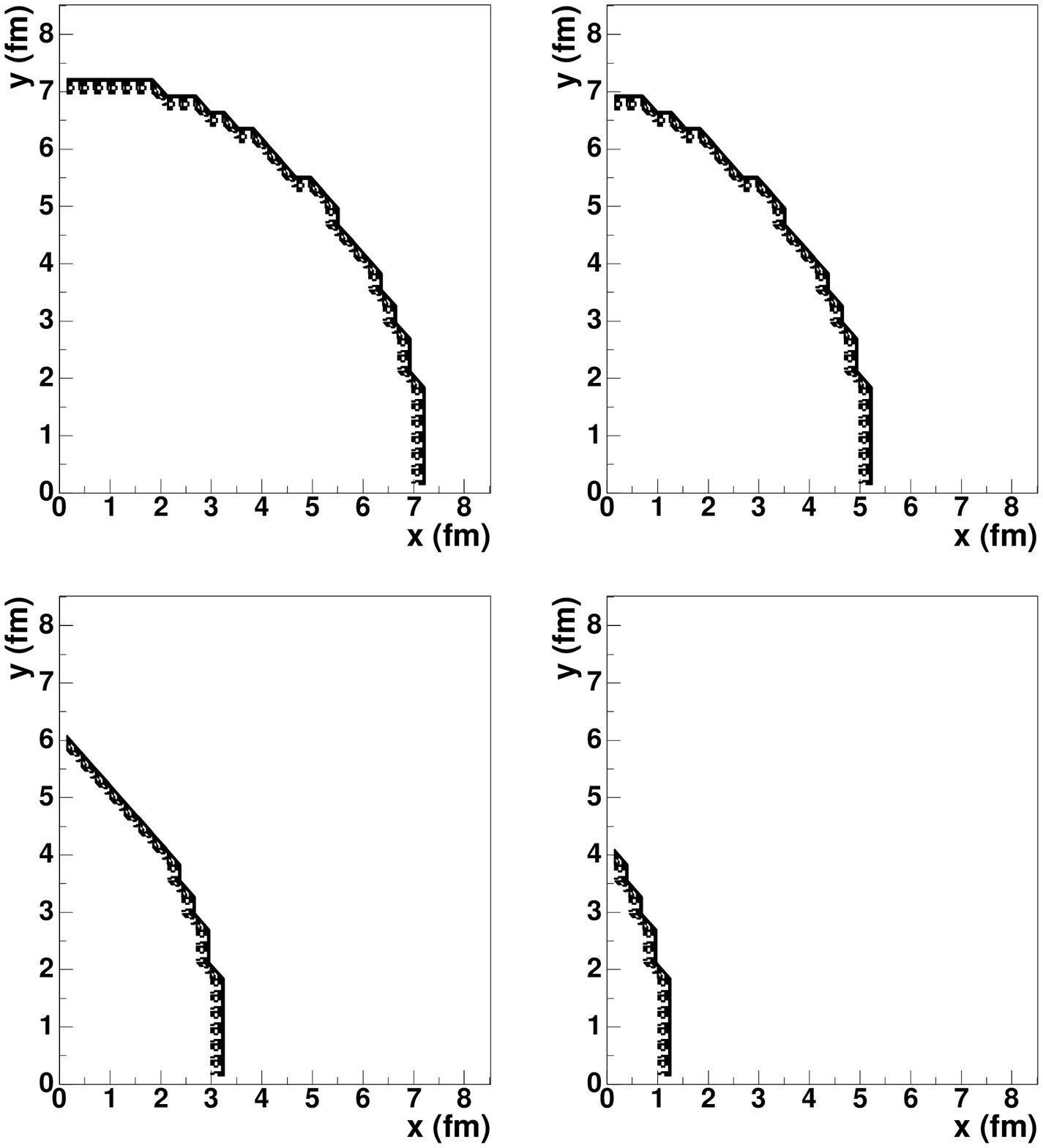}}
\caption{Two dimensional transverse density profile as 
  a function of (x,y) (see Fig~\ref{fig:coordsystem}) corresponding to
  the region of overlap in Fig~\ref{fig:coordsystem}, for Au+Au
  collisions at impact parameters b=0, 4, 8 and 12 fm.}
\label{fig:densitytwod}
\end{figure}


We first present the transverse density profiles graphically, and then
calculate moments to compare the distributions quantitatively. Figures
\ref{fig:densitywn} to \ref{fig:densitytwod} show the transverse
density profiles for Au+Au collisions as linear contour plots in one
quandrant of the coordinate system defined in
Fig.~\ref{fig:coordsystem}, for the four weighting functions at
different impact parameters. The Wounded Nucleon and Binary Collision
profiles have very similar shape, with the Binary Collision
distribution falling off slightly faster from the origin.  The less
realistic Hard Sphere and Two Dimensional calculations exhibit larger
aspect ratios for noncentral collisions and unphysical sharp declines
in the density at the boundaries of the overlap region.

\begin{figure}
\resizebox{3in}{!}{\includegraphics{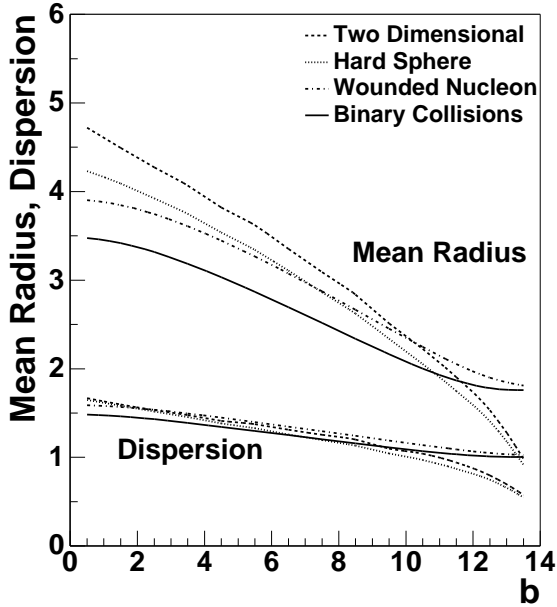}}
\caption{Mean and dispersion of radius of the interaction region as a
  function of impact parameter for Au+Au collisions, for the four
  weighting functions. The units of both axes are fm.}
\label{fig:radius}
\end{figure}

Figure \ref{fig:radius} shows the mean transverse radius
(\smean\ for Fig.~\ref{fig:coordsystem}) and its dispersion
($\sqrt{<\!\absvecs^2\!>-\smean^2}$) for the four weighting functions, as a
function of impact parameter for Au+Au collisions. As is seen in the
density profiles themselves, the Two Dimensional and Hard Sphere
functions give larger mean radii than the Wounded Nucleon and Binary
Collision functions. The dispersion in the radius is similar for all
four functions, and is a weak function of impact parameter. The values
at large impact parameter are dominated by the treatment of the
nuclear surface.

Note that in the case of Binary Collision weighting, the mean (and
median) radius is about 3.5 fm for the most central collisions. In
other words, about one half of all produced e.g. \jpsi\ or jets are
generated farther than 3.5 fm. in the transverse plane from the center
of the reaction zone in central collisions, with a distribution in
radius which has a half-width of about 1.5 fm. Rather few of the \jpsi\ or
jets are produced near the center of the reaction, due simply to the
geometry of nucleus-nucleus collisions.

\begin{figure}
\resizebox{3in}{!}{\includegraphics{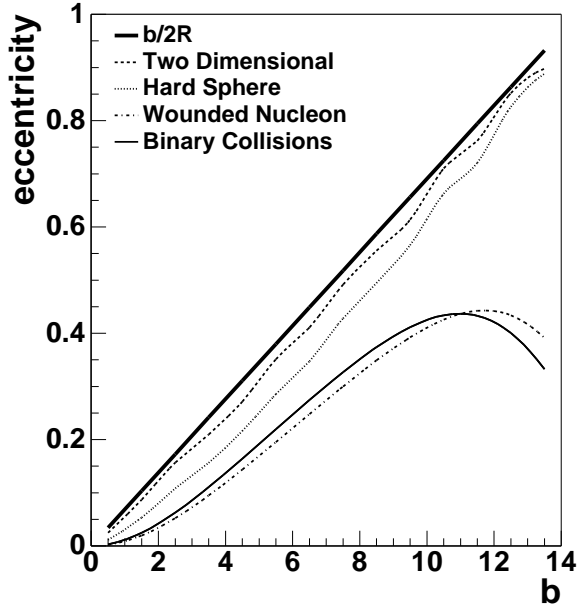}}
\caption{Eccentricity (Eq.~\ref{eq:epsilon}) of the interaction 
  region as a function of impact parameter for Au+Au collisions, for
  the four weighting functions.}
\label{fig:eccentricity}
\end{figure}

Figure \ref{fig:eccentricity} shows the eccentricity
(Eq.~\ref{eq:epsilon}) of the interaction region as a function of
impact parameter for Au+Au collisions, for the four weighting
functions. Also shown is the approximation b/(2\RsubA) from Eq.
\ref{eq:epsilon}. The eccentricity of the Wounded Nucleon and Binary
Collision models are similar and significantly smaller than those of
the other models. Parametrization of $\epsilon$ in the Wounded Nucleon
Model and its dependence upon \sigmaNN\ is given in the next section.

\section{Parametrization of $\epsilon$ and $S$ in the Wounded Nucleon Model}

In this section we give a more detailed discussion of the
characterization of the geometry of the interaction region in the
Wounded Nucleon Model. The elliptic anisotropy of the overlap region
$\epsilon$, defined in Eq.~\ref{eq:epsilon}, is shown as a function of
impact parameter in Fig. \ref{fig:eccentricity} for all four models
for Au+Au collisions. $\epsilon$ is parametrized for the Wounded
Nucleon Model (\sigmaNN=30 mb) as
\begin{equation}
\epsilon(\mathrm{b}) = -0.0469x + 2.754x^2 - 4.797x^3 + 4.852x^4 - 2.492x^5
\label{eq:epsilonparam}
\end{equation}
where $x=\mathrm{b}/(2\mathrm{R})$ and $2\mathrm{R}=15$ fm. for Pb+Pb
collisions and $2\mathrm{R}=14.7$ fm. for Au+Au collisions.

Increase of \sigmaNN\ to 40 mb decreases $\epsilon$ by 13\% at b=2.5
fm and 4\% at b=10.5 fm, with the change monotonic in b. The decrease
of $\epsilon$ with increasing \sigmaNN\ can be understood by noting
that the value of $\epsilon$ is dominated by surface effects.
Increasing \sigmaNN\ effectively extends the surface to larger radii,
with the consequence that the shape of the interaction region in the
transverse projection is less eccentric than for \sigmaNN=30 mb at the
same impact parameter.

For reference, we also include a parametrization in the same model of
the impact parameter dependence of the quantity $S=\pi{R_x}{R_y}$
\cite{heiselberg}. Here, $R_x=\sqrt{<\!x^2\!>}$,
$R_y=\sqrt{<\!y^2\!>}$, and $<\!\ldots\!>$ denotes the spatial average
used in Equation \ref{eq:epsilon}. In \cite{heiselberg}, $S$ is used
to calculate the particle density in the overlap region. The
parametrization is
\begin{equation}
S(\mathrm{b})/(\pi\mathrm{R}^2) = 0.164 + 0.0141x - 0.684x^2 + 1.026x^3 - 0.763x^4 + 0.284x^5
\label{eq:sparam}
\end{equation}
where, again, $x=\mathrm{b}/(2\mathrm{R})$ and $2\mathrm{R}=15$ fm. for Pb+Pb
collisions and $2\mathrm{R}=14.7$ fm. for Au+Au collisions.


\begin{table}[thb]
\begin{center}
\begin{tabular}{ | c | c | c | c | c | }
\hline
b (fm) & \# part & \# part & \# part (HS) & \# BC ($\mu\mathrm{barn}^{-1}$) \\
       & (WN, \sigmaNN=30 mb) & (WN, \sigmaNN=40 mb) & &  \\
\hline
\hline
 0.5 & 368 & 376 & 389 & $2.8\cdot10^{-2}$ \\
\hline
 1.5 & 355 & 364 & 366 & $2.7\cdot10^{-2}$ \\
\hline
 2.5 & 331 & 341 & 335 & $2.5\cdot10^{-2}$ \\
\hline
 3.5 & 298 & 310 & 302 & $2.2\cdot10^{-2}$ \\
\hline
 4.5 & 262 & 274 & 263 & $1.8\cdot10^{-2}$ \\
\hline
 5.5 & 223 & 234 & 226 & $1.5\cdot10^{-2}$ \\
\hline
 6.5 & 183 & 193 & 187 & $1.2\cdot10^{-2}$ \\
\hline
 7.5 & 144 & 155 & 148 & $8.5\cdot10^{-3}$ \\
\hline
 8.5 & 108 & 117 & 115 & $5.8\cdot10^{-3}$ \\
\hline
 9.5 & 77 & 84 & 82 & $3.7\cdot10^{-3}$ \\
\hline
 10.5 & 49 & 55 & 54 & $2.1\cdot10^{-3}$ \\
\hline
 11.5 & 28 & 32 & 32 & $1.0\cdot10^{-3}$ \\
\hline
 12.5 & 13 & 16 & 15 & $4.0\cdot10^{-4}$ \\
\hline
 13.5 & 5 & 6 & 4 & $1.3\cdot10^{-4}$ \\
\hline
\hline
\end{tabular}
\end{center}
\caption{Impact parameter dependence for Au+Au collisions of the
  number per nuclear collision of nucleon participants in the Wounded Nucleon model (columns
  2 and 3) and colliding Hard Spheres (column 4), and the number of
  binary scatterings with cross section 1 $\mu$barn (column 5).}
\label{tab:ncollision197}
\end{table}

\begin{table}[thb]
\begin{center}
\begin{tabular}{ | c | c | c | c | c | }
\hline
b (fm) & \# part & \# part & \# part (HS) & \# BC ($\mu\mathrm{barn}^{-1}$) \\
       & (WN, \sigmaNN=30 mb) & (WN, \sigmaNN=40 mb) & &  \\
\hline
\hline
 0.5 & 388 & 397 & 410 & $3.0\cdot10^{-2}$ \\
\hline
 1.5 & 375 & 385 & 388 & $2.9\cdot10^{-2}$ \\
\hline
 2.5 & 351 & 362 & 358 & $2.6\cdot10^{-2}$ \\
\hline
 3.5 & 318 & 330 & 322 & $2.3\cdot10^{-2}$ \\
\hline
 4.5 & 281 & 293 & 286 & $2.0\cdot10^{-2}$ \\
\hline
 5.5 & 241 & 252 & 246 & $1.6\cdot10^{-2}$ \\
\hline
 6.5 & 199 & 211 & 206 & $1.3\cdot10^{-2}$ \\
\hline
 7.5 & 159 & 170 & 169 & $9.5\cdot10^{-3}$ \\
\hline
 8.5 & 121 & 131 & 132 & $6.7\cdot10^{-3}$ \\
\hline
 9.5 & 87 & 95 & 99 & $4.3\cdot10^{-3}$ \\
\hline
 10.5 & 58 & 64 & 69 & $2.5\cdot10^{-3}$ \\
\hline
 11.5 & 34 & 39 & 44 & $1.3\cdot10^{-3}$ \\
\hline
 12.5 & 17 & 20 & 23 & $5.5\cdot10^{-4}$ \\
\hline
 13.5 & 7 & 9 & 8 & $1.9\cdot10^{-4}$ \\
\hline
\hline
\end{tabular}
\end{center}
\caption{Impact parameter dependence for Pb+Pb collisions of the
  number per nuclear collision of nucleon participants in the Wounded Nucleon model (columns
  2 and 3) and colliding Hard Spheres (column 4), and the number of
  binary scatterings with cross section 1 $\mu$barn (column 5).}
\label{tab:ncollision208}
\end{table}
\section{Number of Participants and Binary Collisions}

We conclude with a tabulation of the number of participants and binary
collisions as a function of impact parameter. The distributions in
Figures \ref{fig:densitywn} to
\ref{fig:densityhs} can be integrated to calculate the total number of
nucleon participants or binary collisions per nuclear collision.
These are presented in Tables
\ref{tab:ncollision197} and \ref{tab:ncollision208} for Au+Au and
Pb+Pb collisions. Columns 2 and 3 give the number of Wounded Nucleons
(Equation \ref{eq:woundednucleon}) with \sigmaNN=30 and 40 mb. Column
4 gives the fraction of volume of the Hard Spheres that interact,
normalized to the total number of incoming nucleons. Column 5 gives
the number of binary collisions for an interaction cross section of 1
$\mu\mathrm{barn}$ (the rate for other small cross sections is
obtained by linear scaling of this number).

Tables \ref{tab:ncollision197} and \ref{tab:ncollision208} show only a
weak dependence of the number of participants on \sigmaNN, and good
agreement between the number of participants calculated with the Hard
Sphere and Wounded Nucleon models. The latter point is in sharp
contrast to the strong difference seen between the models for
$\epsilon$ (Figure \ref{fig:eccentricity}), and is due to the fact
that the number of participants is dominated by the bulk volume,
whereas $\epsilon$ is dominated by the surface overlap.


\section{Acknowledgements}

We thank Art Poskanzer, Raimond Snellings, and Sergei Voloshin for valuable discussions.


\end{document}